\documentclass[twocolumn,aps,pra,superscriptaddress,floatfix]{revtex4}

\usepackage{graphicx}
\usepackage{amsmath}
\usepackage{amssymb}

\begin{document}
\title{Continuous variable quantum key distribution with modulated entangled states}

\author{Lars S. Madsen}
\email{lsma@fysik.dtu.dk}
\affiliation{Department of Physics, Technical University of Denmark, Fysikvej, 2800 Kongens Lyngby, Denmark}

\author{Vladyslav C. Usenko}
\affiliation{Department of Optics, Palack\' y University, 17. listopadu 12,  771~46 Olomouc, Czech Republic}
\affiliation{Bogolyubov Institute for Theoretical Physics of National Academy of Sciences,
Metrolohichna st. 14-b, 03680, Kiev, Ukraine}

\author{Mikael Lassen}
\affiliation{Department of Physics, Technical University of Denmark, Fysikvej, 2800 Kongens Lyngby, Denmark}

\author{Radim Filip}
\affiliation{Department of Optics, Palack\' y University, 17. listopadu 12,  771~46 Olomouc, Czech Republic}

\author{Ulrik L. Andersen}
\affiliation{Department of Physics, Technical University of Denmark, Fysikvej, 2800 Kongens Lyngby, Denmark}

\date{\today}


\maketitle

\textbf{
Quantum key distribution (QKD) enables two remote parties to grow a shared key which they can use for unconditionally secure communication over a certain distance. The maximal distance depends on the loss and the excess noise of the connecting quantum channel. Several QKD schemes based on coherent states and continuous variable (CV) measurements are resilient to high loss in the channel, but are strongly affected by small amounts of channel excess noise. Here we propose and experimentally address a CV-QKD protocol that uses modulated fragile entangled states of light to greatly enhance the robustness to channel noise. We experimentally demonstrate that the resulting QKD protocol can tolerate more noise than the benchmark set by the ideal CV coherent state protocol. Our scheme represents a very promising avenue for extending the distance for which secure communication is possible.}


\section*{Introduction}

 There is a tremendous demand for secure communication of data in commerce, finance and government affairs. Unconditional security is promised by the use of a one-time pad strategy where two parties, Alice and Bob, share a pre-established secret key which they use for encoding and decoding the message. The confidentiality of the communication therefore falls back on the generation of a secret key between Alice and Bob~\cite{qkd2,rev}. Such a key can be generated with quantum key distribution (QKD) which was first proposed by Bennett and Brassard in 1984 (BB84) for single photons and discrete variable measurements~\cite{bb84}. The technology has later been extended to also include coherent states and continuous variable (CV) measurements known as CV-QKD~\cite{coh1,exp1,exp3,silber1,piran,exp4,nonswitching,twoway,raul,colclon}.

A generic CV-QKD protocol between two trusted parties is initiated by Alice, who prepares a distribution of Gaussian quantum states of light, e.g., coherent~\cite{coh1,exp1,exp3,exp4,silber1,nonswitching,colclon,piran} or squeezed/entangled states~\cite{sq0,sq1,sq2,sq3,eprexp,raul}. Alice transmits the states through a quantum channel to Bob, who performs measurements on the continuous quadrature components of the light field using either a homodyne detector~\cite{coh1,exp1,exp3, colclon} or a heterodyne detector~\cite{nonswitching}, thus measuring conjugate quadratures either randomly or simultaneously, respectively. This results in a set of data that is partially correlated with Alice's data set, which she obtained in the process of preparing the distribution of quantum states. Alternatively, a two-way quantum communication scheme can be formulated~\cite{twoway}, but we will restrict our discussion to one-way quantum communication. To estimate the secrecy of the transmission, Alice and Bob compare a subset of their data using classical communication. Provided that the security threshold for channel loss and excess noise has not been crossed, the resulting set of raw data can then be mapped onto a shared secret key using classical reconciliation and  post-processing techniques~\cite{recon,key,exp3}.

There are two major hurdles in CV-QKD that limit the distance for secure communication. The first is the presence of excess noise, combined with high losses in the optical channel~\cite{coh1,exp1,exp3,colclon,nonswitching}, and the second is the limited classical reconciliation efficiency. For example, in the realistic CV-QKD scheme based on coherent states, the maximal secure distance is in theory limited to around 140 km if the channel excess noise is 4\% of vacuum noise, the loss is 0.2 dB/km and the post-processing efficiency is 96.9\%~\cite{key}. To enlarge the secure distance, one obvious strategy is to reduce the channel loss and noise, and to increase the post-processing efficiency. However, present CV-QKD systems are already working with state-of-the-art optical channels and the post-processing efficiency is also reaching its limit. Therefore, to go beyond the currently achievable distances, a fundamentally different approach must be followed.   

In this Article we propose and, as a proof-of-principle, experimentally demonstrate a CV-QKD protocol based on 
 entangled states of light that is more tolerant to channel excess noise, channel loss and limited post-processing efficiency than the coherent state based protocols. 
Explicitly, using only 3.5 dB of impure and modulated two-mode squeezing, we demonstrate the generation of a secret raw key between two parties connected by a noisy and lossy channel - a channel which cannot be used for secure communication based on any coherent state protocol. A full scale implementation of our proposed protocol has the capability to significantly boost the robustness and distance for secure communication using available technology and a feasible, that is impure, squeezed light source.

\begin{figure}[h]

\begin{center}
\includegraphics[width=3.4in]{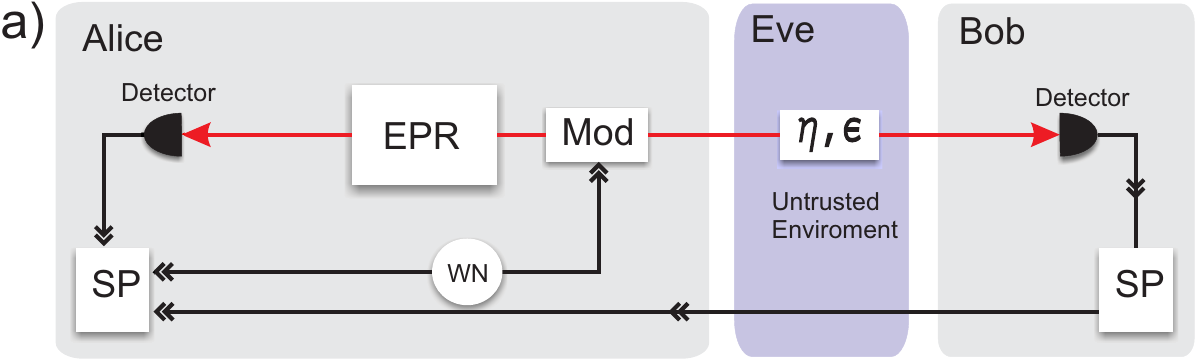}
\includegraphics[width=3.4in]{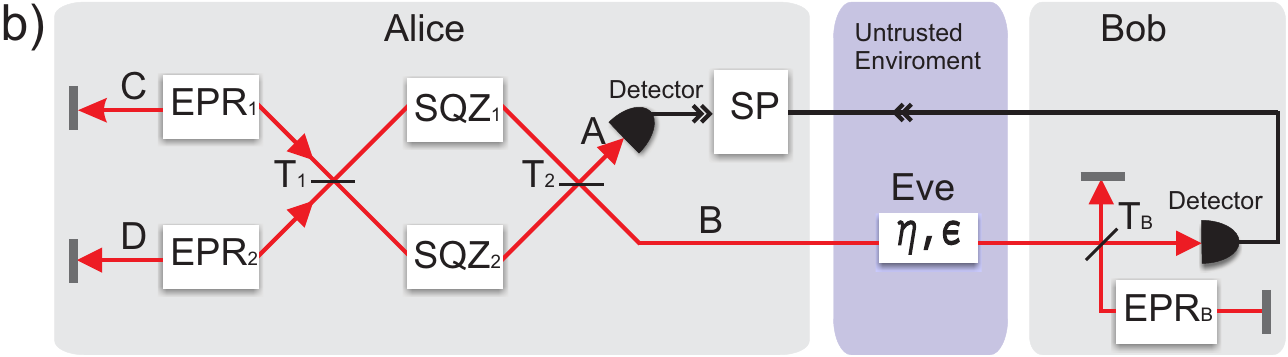}
\caption{\textbf{Conceptual diagram and purification of our QKD protocol.} a) Alice prepares a conditionally squeezed state by randomly measuring the amplitude or the phase quadratures of one mode of an EPR state using a homodyne detector. The conditionally squeezed state is modulated further by a modulator (Mod) fed with a Gaussian white noise (WN) source controlled by Alice. The homodyne data and the white noise data are stored for signal processing (SP) and the gain between them is optimized. The modulated conditionally squeezed state is transmitted through an untrusted quantum channel where Eve is allowed to perform any attack that mimics the channel transmission $\eta$ and the channel excess noise $\epsilon$. After the channel Bob performs quadrature measurements using a homodyne detector and the classical post-processing can begin. b) Purification scheme for an arbitrary Gaussian QKD protocol.  Two quadrature squeezers (SQZ$_1$ and SQZ$_2$) are placed inside a Mach-Zehnder interferometer with beam splitters of transmittances T$_1$ and T$_2$ and fed with modes from two independent EPR sources (EPR$_1$ and EPR$_2$) . The resulting 4-mode state (A,B,C and D) is pure, while the six free parameters can be set so that the two modes A and B can simulate any Gaussian two-mode state (up to a local unitary transformation) including the states produced in our experimental setup. One mode of the state is measured by an ideal detector at Alice while the other travels through the channel which has transmission $\eta$ and excess noise $\epsilon$. Finally Bob's noisy detection is purified by placing a beamsplitter with an EPR input and a transmission mimicking his electronic noise, detection efficiency and the noise he adds to his data, before an ideal detector.}
\label{theosetup}
\end{center}
\end{figure}

\section*{Results}
{\bf Our protocol.} In our scheme we prepare a Gaussian distribution of squeezed states using the method illustrated in Fig.~\ref{theosetup} a). Alice prepares a Gaussian entangled state, known as an Einstein-Podolsky-Rosen (EPR) state, 
and measures one of the modes using a homodyne detector that randomly detects the amplitude or the phase quadrature. This measurement projects the EPR state onto a Gaussian distribution of conditionally squeezed states \cite{epr}. Such alphabet of squeezed states could in principle be used to outperform the coherent state protocol under the condition of pure and strong entanglement~\cite{raul}. To release these stringent requirements, we propose to enlarge the Gaussian distribution in phase space by a controlled modulation using two random and independent Gaussian variables. The final Gaussian distribution of states is then transmitted through a potentially lossy and noisy quantum channel the action of which may be ascribed to an eavesdropper (Eve). Finally, the states are measured by Bob who randomly measures one of the two conjugate continuous quadratures using homodyne detection. 

After the transmission, Alice holds two sets of data: one set obtained from the homodyne measurements, $\{x_{HD}\}$, and one from the Gaussian modulation, $\{x_{M}\}$. To maximize the performance of the protocol, we suggest to weight the homodyne data with a gain factor, $g \in \left[0,1\right]$, and subsequently add the two sets to yield the optimized set: $x_{M}+gx_{HD}$. The optimal gain factor depends on the strength and the purity of the EPR state. 
In the limit of no squeezing, 
 Alice only keeps the data from the Gaussian modulation and thus the protocol reduces to the standard coherent state protocol~\cite{coh1,exp1}. On the other hand, for very high antisqueezing, $g=1$ and the values of the resulting data set are equally constructed from the two subsets. However, in a real life scenario squeezing is limited and thus an intermediate gain will optimize  the generation of a secret key. The rest of the classical part of the protocol follows the common recipe of the generic protocols.

{\bf Secret Key.} To prove the security of our CV-QKD protocol, it is sufficient to consider the class of collective attacks as they have been shown to be optimal under certain symmetries of the protocol \cite{extremality,proof1,proof2,finetti}. 
For collective attacks and using the classical technique of reverse reconciliation~\cite{exp1}, the achievable key rate (in the asymptotic limit of an infinitely long raw key) is given by 
\begin{equation}
\label{keyrate}
I= \beta I_{AB}-\chi_{BE},
\end{equation}
where $I_{AB}$ is the Shannon mutual information between the data of Alice and Bob, $\chi_{BE}$ is the Holevo bound on the information available between Bob and Eve and $\beta$ is the post-processing efficiency. 

To use this general framework for the calculations of the key rate, it is necessary to define a theoretical preparation scheme~\cite{vlad1,fred,equiv} where the actual two-mode state of Alice and Bob is part of a pure multi-mode state~\cite{braunstein2005}. This purification is generated by using pure sources and accounting for all modes. For example, the standard prepare-and-measure coherent state or squeezed state protocols are often treated theoretically by an equivalent entanglement based protocol. In the present work, the actual setup is already partially entanglement based however the resulting states are impure due to imperfections in the squeezing sources and the imposed modulations. We therefore define a theoretical preparation scheme (see Fig. \ref{theosetup} b)) that can generate any two-mode Gaussian state between Alice and Bob as part of a pure four-mode state (see Methods).

\begin{figure}
\begin{center}
\includegraphics[width=0.4\textwidth]{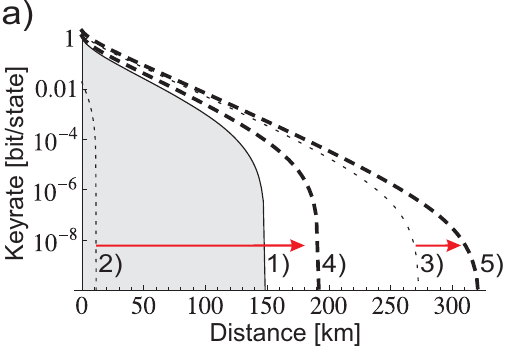}
\includegraphics[width=0.4\textwidth]{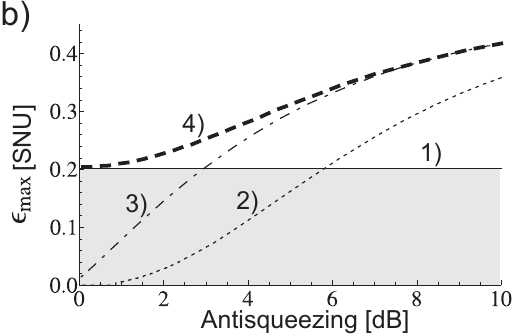}
\caption{\textbf{Theoretical comparison between the performance of different CV-QKD protocols.} a) Secret key rate as a function of distance (corresponding to a loss of 0.2 dB per km) for a fixed excess noise of 0.1 SNU.  1) Ideal coherent state protocol with 100 SNU modulation. The shaded region illustrates the regime accessible with coherent state protocols.  2) and 3) Squeezed state protocol with 3 dB and 10 dB squeezing respectively (without additional modulation). 4) and 5) Our proposed protocol with 3 dB and 10 dB of squeezing respectively and 100 SNU modulation. The red arrows indicate the improvement of our proposed protocol compared with the squeezed state protocol with no modulation. For all protocols, the added noise to Bob's data is optimized and $\beta=1$.
b) Maximal tolerable channel noise versus the initial antisqueezed variance. The channel loss is set to 10 dB (corresponding to a distance of 50 km). 1) Ideal coherent state protocol with asymptotically large modulation. The shaded region illustrates the regime accessible with coherent state QKD. 2) Squeezed state protocol without additional modulation. 3) New combined squeezed state protocol with 100 SNU of coherent modulation without the gain. This is also the performance obtained for highly impure squeezed states. 4) Our proposed optimized protocol with 100 SNU coherent modulation and optimized gain factor. For all protocols, the added noise to Bob's data is optimized and $\beta=1$.}
\label{theory1}
\end{center}
\end{figure}

To reach large distances for which key distribution can be attained, one must develop a protocol that maximizes the key rate in Eq. (\ref{keyrate}). We start by considering the standard coherent state protocol assuming unity post-processing efficiency. It is known that the key rate for this protocol can be maximized by using an infinitely large Gaussian modulation and by adding white noise to Bob's data, which are obtained by homodyne detection~\cite{raul}. 
Using the above mentioned analysis for collective attacks, we calculate the secure key rate and the maximally tolerable excess noise, and the results are illustrated by the solid curves in Fig. \ref{theory1} a) and b) . These two curves represent the coherent state benchmarks. Employing the squeezed state protocol suggested in ref.~\cite{raul}, these benchmarks can be beaten but only for strongly squeezed states, that is, 5.6 dB noise suppression below the shot noise limit as illustrated by the dotted curves in Fig. \ref{theory1} a) and b). 

Now by considering our protocol (Fig. \ref{theosetup} a)), the key rate and the tolerable excess noise are increased even further, as shown by the bold dashed lines in Fig. \ref{theory1}. Comparing the previous squeezed state protocol \cite{raul} with ours, we see that the maximal secure distance attainable for 3dB squeezed states is increased by a factor of about 19 and the required squeezing for surpassing the coherent state protocol is lowered from 5.6 dB to $0$ dB for pure two-mode squeezed states. For highly impure two-mode squeezed states we need 3 dB of two-mode squeezing as shown by the dot-dashed curve. Our protocol thus has the remarkable property that any conditionally squeezed state improves the performance beyond the optimized coherent state protocol. Moreover, another important feature of our protocol is that the squeezed states need not be pure; arbitrary mixedness can be tolerated as long as the state is conditionally squeezed. Therefore, the main resource for increased performance is conditional squeezing. We note however that the performance saturates for high degrees of squeezing (see. Fig.\ref{theory1} b).

\begin{figure}[h]
\begin{center}
\includegraphics[width=3.4in]{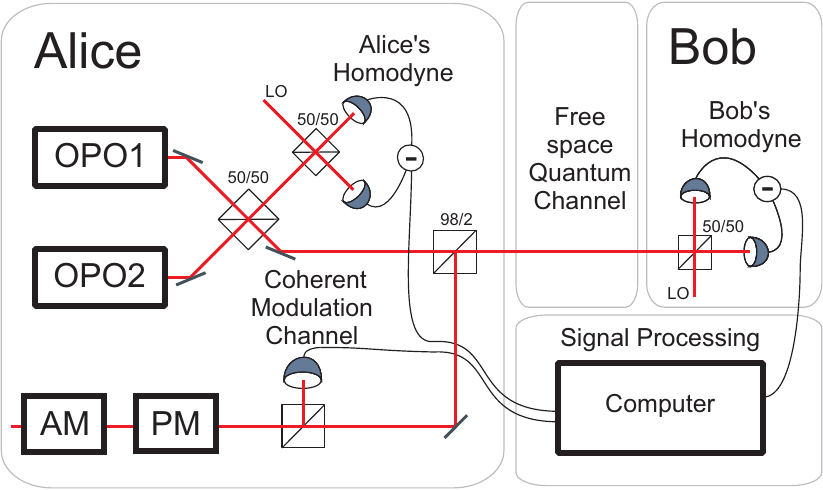}
\caption{\textbf{Experimental setup.} A squeezed state is generated in each of the two optical parametric oscillators (OPOs), operated below threshold. The squeezed states interfere at a beam splitter to form a two-mode squeezed state.  The amplitude or the phase quadrature of the one mode is measured by Alice's homodyne detector. The other mode is carefully phase locked to a coherently modulated auxiliary mode. The auxiliary state is generated using a phase modulator (PM) and an amplitude modulator (AM), each of which are driven by a white noise generator. Alice acquires information about the modulation by measuring a part of the auxiliary state. The coherent state is purified by the highly asymmetric beamsplitter all of this in order to minimize the harmful preparation noise \cite{vlad1}. The modulated and conditionally squeezed state is transmitted through the channel to Bob's homodyne detector where one of the conjugate quadratures is measured. The resulting measurement outcomes are fed via a fast AD card to a computer for signal processing.}
\label{test}
\end{center}
\end{figure}

{\bf Experimental setup and results.} The experimental setup is sketched in Fig. \ref{test}. We start by generating EPR entanglement between two modes of light~\cite{epr}. The quadrature of one of the EPR modes is measured by means of high efficiency homodyne detection at Alice's station. In addition to the measurement, we induce a random but known coherent modulation to the second EPR mode. After the channel we measure the second mode using a high efficiency homodyne detector at Bob's station, to access either the amplitude quadrature or the phase quadrature. We use laser light at 1064 nm for the seeds and local oscillators and 532 nm for the pump for the optical parametric oscillators (OPOs). The EPR state has $3.5$ dB $\pm0.2$ dB of two-mode squeezing and 8.2 dB $\pm0.2$ dB of antisqueezing and the coherent modulation depth is sequentially varied between 0 and 15 dB. All measurements are performed at the sideband frequency of 4.9 MHz with a bandwidth of 90 kHz.  The system is initially calibrated to ensure that the correlation between conjugate quadratures of Alice is negligible and that the quadratures of Alice are in phase with Bob. 

The experiment was carried out sequentially for conjugate quadratures; the amplitude quadrature was first conditionally squeezed, displaced and measured, and the procedure was then repeated for the phase quadrature.  We add the homodyne and modulation data with an optimized gain. For technical reasons, we multiply the modulation data with the inverse of the gain factor. This results in two large data strings, one for Alice and one for Bob, which are strongly correlated as shown  for three different modulation depths in Fig. \ref{data1} a)-c ). The correlations arise partly from the quadrature entanglement and partly from the coherent modulation. From the correlated data we compute the covariance matrices as illustrated in Fig. \ref{data1} d)-f), from which we can estimate the security limits for our system (see Methods). 

\section*{Discussion}

In the laboratory the channel transmission is $95\%$ and the excess noise can be continuously varied by unbalancing the generated data at Alice and Bob. As an example we set the excess noise to 0.45 shot noise units (SNU) and the total modulation depth to 23.4 SNU. For these settings of the experiment we generate a raw key with a rate of $0.004 \pm 0.001$ bit per state. We note that neither the coherent state based protocols nor the standard squeezed state based protocol (with $3.5$ dB squeezing) could have generated a key in such a channel.

\begin{figure}[h!]
\begin{center}
\includegraphics[width=0.6\textwidth]{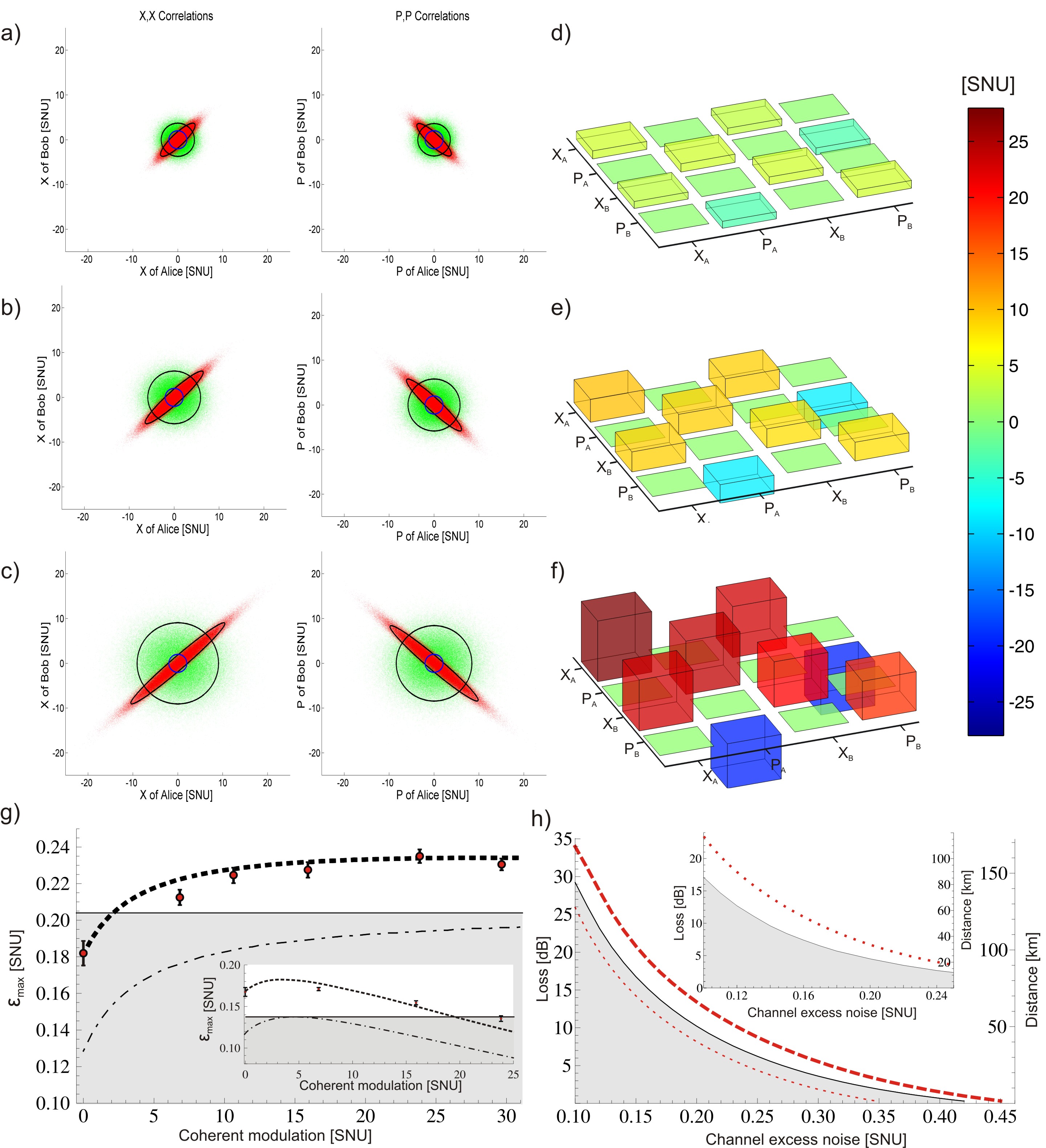}
\caption{\textbf{QKD measurement results.} a-c) The red points represent the normalized and weighted quadrature data of Alice and Bob for three different depths of coherent modulation. In the left and the right the amplitude and phase quadrature correlations are shown, respectively. The extra modulation added is 0 SNU in a), 3.6 SNU in b) and 23.8 SNU in c), which comes on top of 3.6 SNU modulation from the EPR source. These correlated data are contrasted with a set of uncorrelated data points (green) with identical total energy.  The solid black ellipse and outer circle correspond to two standard deviations of these two respectively. The inner blue circle is two standard deviations of shot noise. d-f) Illustrations of the covariance matrices for the three modulation depths. g) Tolerable excess noise as a function of the modulation depth for a simulated transmission of $\eta=0.1$. The uncertainty represents the actual measurements with compensation for $85\pm 5\%$ quantum detection efficiency as the dominating source of uncertainty. The theoretical estimates for the ideal coherent state protocol and our protocol with the experimental parameters are given by the solid and dashed curves, respectively. For comparison, we also include the performance of the coherent state protocol with an energy identical to the energy of the two-mode squeezing protocol (dot dashed curve). The shaded region illustrates the regime accessible with coherent state QKD. The inset is the same as the main figure but with limited post-processing efficiency $\beta=98\%$. h) The maximally tolerable loss as a function of the channel excess noise. The thick dashed line is for the data set presented in c) and f), the solid curve is for the ideal coherent state protocol. The dotted curve corresponds to the two-mode squeezing protocol without additional modulation presented in matrix a) and d). The shaded region illustrates the regime accessible with coherent state QKD. The distance scale assumes 0.2 dB loss per km. The insert is the same as the main figure, the dotted curve being the data set presented in a) and d) but with limited post-processing efficiency $\beta=98\%$.}
\label{data1}
\end{center}
\end{figure}

We now investigate the security performance of our protocol in longer channels based on the experimentally measured covariance matrices assuming perfect post-processing and channel estimation. The matrices are used in a model that includes the trusted losses and noise sources of the detectors, and in which arbitrary channel loss and excess noise can be simulated. As an example we assume a channel transmission of 10\% and find the tolerable excess noise for six different realizations of the covariance matrices illustrated in Fig.~\ref{data1} g). Finally, in  Fig.~\ref{data1} h) we plot the maximum distance and loss as a function of the tolerable noise associated with the experimentally realized covariance matrix in  Fig.~\ref{data1} f). We clearly see that by combining squeezed states with coherent modulation we beat the performance of any coherent state protocol (limited to the shaded region). 
The supremacy of the squeezed state protocol relative to the coherent state protocol is best seen by equalizing the amount of energy used in the two protocols. In this case, the states entering the channel are identical and Eve cannot tell the difference between the two protocols. The relative improvement is illustrated by the dot-dashed curve relative to the dashed curve in Fig.~\ref{data1} g).  



In the analysis above we have assumed perfect classical post-processing (corresponding to $\beta=1$). However, the key rate of real world implementations of CV-QKD is currently limited by the inefficiency of the yet developed classical error codes~\cite{key,beta}. We have therefore considered the effect of imperfect post-processing on our protocol corresponding to a security analysis with $\beta<1$. The results of this analysis for an optimistic $\beta=0.98$ are illustrated in the insets of Fig.~\ref{data1} g) and Fig.~\ref{data1} h), while a more pessimistic $\beta=0.95$ can be seen in supplementary Figure S1, and further discussed in the supplementary discussion. Finally we fix the channel transmission to 10\% and plot the tolerable excess noise as a function of the post-processing efficiency in supplementary Figure S2.
It is evident from the figures that despite inefficient post-processing, our squeezed state protocol remains superior to the coherent state protocol, and remarkably, it is seen that the relative improvement has increased. We also note that the optimal performance occurs for a finite modulation depth. 
If the beta factor becomes very low, a different squeezed state protocol must be considered~\cite{our}. 

As mentioned above, the performance of our protocol is partially parametrized by the degree of conditional squeezing. In our realization, the conditionally squeezed states were prepared by measuring one mode of a two-mode squeezed state. However, alternatively, the conditionally squeezed state could have been prepared directly by modulating a single-mode squeezed states beyond the antisqueezing. The direct benefit of such a strategy is that the degree of conditional squeezing is identical to the degree of the single mode squeezed state. 
Protocols with single mode squeezed states will be studied in future work.

We have introduced and experimentally addressed a QKD protocol based on continuous variable squeezed states of light. Interestingly, we find that the key rate as well as the robustness against channel noise is improved for any degree of conditional squeezing compared to the idealized and optimized coherent state protocol. The requirement on the purity of the squeezed states is relaxed, thus rendering the protocol provably secure against any type of attack on the channel. In fact, by using strongly squeezed but also highly impure states, such as those produced via the large bandwidth Kerr effect in standard optical fibers (where up to 6.8 dB of squeezing and 29.6 dB of anti-squeezing has been produced \cite{fiber}), the coherent modulation is left unnecessary. In future commercial implementations, miniaturized waveguide cavities could be used to make on chip squeezed state QKD. 

\section*{Methods}

\subsection{Theory of the protocol}

After obtaining the two-mode covariance matrix, the security analysis of our protocol follows the well established security proofs for the Gaussian CV-QKD protocols, which are based on the extremality of Gaussian states~\cite{extremality} and consequently the optimality of Gaussian collective attacks~\cite{proof1,proof2}. From this, it follows that the lower bound on the key rate as stated in eq.~(\ref{keyrate}), can be obtained from the covariance matrix of the state. Given that the difference in~(\ref{keyrate}) is positive, the classical privacy amplification algorithms can distill a secure key. 

The Shannon mutual information is calculated directly from the variance of Alice's quadrature data, $V_A$, and Alice's variance conditioned on Bob's measurement, $V_{A|B}$, as $I_{AB}=\frac{1}{2} \textrm{log}_2(V_A/V_{A|B})$. The gain factor, weighting the fraction of Alice's homodyne measurement results, is then optimized to maximize the mutual information.

The Holevo quantity is expressed through the von Neumann entropies $S(\cdot)$  as  $\chi_{BE}=S(\rho_{E})-S(\rho_{E}^{x_B})$ in the reverse reconciliation scheme, where  $\rho_{E}$ is the density matrix of Eve's states and $\rho_E^{x_B}$ is the density matrix conditioned by Bob's measurement.
If untrusted noise is present in the system, Eve is assumed to be able to purify the system of Alice and Bob so that $S(\rho_{E})=S(\tilde\rho_{AB})$ and $S(\rho_{E}^{x_B})=S(\tilde\rho_{AB}^{x_B})$ (assuming, with no loss of generality, that the amplitude quadrature is measured by Bob with result $x_B$), where $\tilde\rho_{AB}$ is the pure state shared between Alice and Bob \cite{exp3}. Thus, in order to calculate the Holevo quantity we need to purify the state of Alice and Bob. In the theoretical analysis of the protocol, it is done by explicitly introducing all trusted modes~\cite{raul,vlad1} and constructing the overall covariance matrix of the trusted pure state. The von Neumann entropies are then calculated from the symplectic eigenvalues of the respective covariance matrices (See supplementary methods for details). 

\subsection{Experimental data security analysis}
The measurement and the subsequent data processing (including gain optimization) resulted in a set of covariance matrices for different values of modulation. The security analysis was carried out according to the above described purification method and cross-checked using the entangling cloner method (which is shown to be optimal for the individual attacks in \cite{equiv} and used for the collective attacks in \cite{colclon}). Since the explicit trusted mode structure of the experimentally obtained states was not known, the experimental covariance matrices were purified using the Bloch-Messiah reduction theorem~\cite{braunstein2005} as illustrated in Fig. \ref{theosetup} b). For each of the covariance matrices the 4-mode pure state (ABCD on Fig. \ref{theosetup} b)) was constructed and used to calculate the Holevo quantity and the resulting lower bound on secure key rate.

In parallel, the theoretical estimation of the expected protocol performance was calculated from the experimental parameters of our setup. The covariance matrices were constructed from the experimentally measured EPR states. All the respective transmittances and efficiencies were applied in both EPR modes (as beamsplitter transformations) and the coherent modulation was added atop. The resulting covariance matrices were tested against given channel transmittance and excess noise.

Furthermore we have simulated the accuracy in estimating the channel as a function of the size of the data blocks. The results are shown in the Supplementary Figure S3) and discussed in the Supplementary discussion.

\subsection{Experimental setup details}
Our light source is an Innolight Diabolo laser producing light at 1064 nm and 532 nm. Both beams are sent through high finesse triangular mode cleaning cavities for temporal and spatial mode cleaning. The mode at 532 nm is used as the pump for the optical parametric oscillators (OPOs) whereas the mode at 1064 nm serves as a lock beam for the OPOs, auxiliary beam for the modulation channel and as local oscillators (LOs) for the homodyne detection systems. The OPOs are 25 cm long bowtie shaped cavities with two curved dichromatic mirrors (with 25 mm radius of curvature and highly reflective at 1064 nm) and two plane mirrors; one with a highly reflective coating and another one with a reflectivity of $8\%$ for OPO1 and $10\%$ for OPO2. This yields cavity bandwidths of around 21 MHz and 24 MHz. The nonlinear crystals inside the OPOs are temperature controlled type I periodically poled KTP crystals which are phase matched for down conversion at the specified wavelenghts. Each of the crystals is pumped with 170 mW of the 532 nm beam, which means that the OPOs are operating well below threshold. A small part of the squeezed output beam is tapped of and used for locking the phase of the pump to deamplification thereby generating amplitude squeezed sidebands. The states measured directly have $4.9$ dB and $4.1$ dB of shot noise reduction along the squeezed quadratures and $8.3$ dB of increased noise along the conjugate quadratures. When coupling these two squeezed seeds we measure an EPR state with 3.5 dB two-mode squeezing and 8.2 dB of antisqueezing. The total homodyne detection efficiency is $90\% \pm 5\%$ at Alice and $85\% \pm 5\%$ at Bob. The signal is mixed down and low passed with a 90 kHz filter before it is digitalized with a sampling rate of 500 kHz. Each data block consists of around 200.000 points.

\subsection*{Acknowledgments}
This work was supported by the Danish Agency for Science Technology and Innovation (no. 274-07-0509). M.L acknowledges the Danish Council for Technology and Production Sciences (no. 10-093584). V.C.U. was supported by the project no. P205/10/P32 of the Grant Agency of Czech Republic, R.F. acknowledges the support from grant P205/12/0577 of GA \v CR.

\subsection*{Author contributions}
L.S.M and M.L. performed the experiments. V.U. and R.F. performed the theoretical calculations. L.S.M. and V.U. performed the data analysis. All authors wrote the manuscript. L.S.M., M.L. and U.L.A performed the project planing and R.F. and U.L.A supervised the project. All authors discussed the results and implications and commented on the manuscript at all stages.

\subsection*{Competing Interests}
The authors declare that they have no competing financial interests.



\begin{thebibliography}{99}


\bibitem{qkd2} Scarani, V. et al., The security of practical quantum key distribution. \textit{Rev. Mod. Phys.} \textbf{81}, 1301-1350 (2009).

\bibitem{rev} Weedbrook, Ch. et al., Gaussian Quantum Information, \textit{Rev. Mod. Phys.} \textbf{84}, 621-669 (2012).

\bibitem{bb84}  Bennet, C. H. \& Brassard, G.  “Quantum Cryptography: Public key distribution and coin tossing”, in Proceedings of the IEEE International Conference on Computers, Systems, and Signal Processing, Bangalore, p. 175 (1984). 


\bibitem{coh1} Grosshans, F. \& Grangier, P. Continuous variable quantum cryptography using coherent states. \textit{Phys. Rev. Lett.} \textbf{88}, 057902 (2002).
\bibitem{exp1} Grosshans, F. et al. Quantum key distribution using gaussian-modulated coherent states. \textit{Nature} \textbf{421}, 238-241 (2003).

\bibitem{exp3} Lodewyck J. et al., Quantum key distribution over 25 km with an all-fiber continuous-variable system. \textit{Phys. Rev. A} \textbf{76}, 042305 (2007).
\bibitem{colclon} Weedbrook, C. , Pirandola, S. ,  Lloyd, S. \&  Ralph, T. C. Quantum cryptography approaching the classical limit. \textit{Phys. Rev. Lett.} {\bf 105}, 110501 (2010).
\bibitem{nonswitching}Weedbrook C. et al., Quantum cryptography without switching. \textit{Phys. Rev. Lett.} \textbf{93} 170504 (2004).
\bibitem{piran} Pirandola, S., Braunstein, S.L. and Lloyd, S., Phys. Rev. Lett. \textbf{101}, 200504 (2008). 
\bibitem{silber1} Silberhorn, C., Ralph, T. C., Lutkenhaus,  N. \& Leuchs, G. Continuous variable quantum cryptography: Beating the 3 dB loss limit \textit{Phys. Rev. Lett.} \textbf{89}, 167901 (2002).
\bibitem{exp4}  Lorenz, S. ,  Korolkova, N. , \& Leuchs, G. Continuous-variable quantum key distribution using polarization encoding and post selection. \textit{Applied Physics B} \textbf{79}, 273-277  (2004).

\bibitem{twoway} Pirandola, S., Mancini, S., Lloyd, S. \& Braunstein,S.L. Continuous-variable quantum cryptography using two-way quantum communication. \textit{Nature Phys.} \textbf{4}, 726-730 (2008).
\bibitem{raul} Garc\'ia-Patr\'on R. \&  Cerf N. J. Continuous-variable quantum key distribution protocols over noisy channels. \textit{Phys. Rev. Lett.} \textbf{102}, 130501 (2009).

\bibitem{sq1} Hillery, M. Quantum cryptography with squeezed states. \textit{Phys. Rev. A} \textbf{61}, 022309 (2000).
\bibitem{sq2} Cerf, N.J. ,  Levy, M. , \& Van Assche G. Quantum distribution of Gaussian keys using squeezed states. \textit{Phys. Rev. A} \textbf{63}, 052311 (2001).
\bibitem{sq0} Ralph, T.C. Continuous variable quantum cryptography \textit{Phys. Rev. A}  \textbf{61}, 010303(R) (1999).
\bibitem{sq3} Gottesman, D. \&  Preskill, Secure quantum key distribution using squeezed states. \textit{Phys. Rev. A} \textbf{63}, 022309 (2001).
\bibitem{eprexp} Su, X., et al., Continuous variable quantum key distribution based on optical entangled states without signal modulation \textit{Europhys. Lett.} {\bf87}, 2000-2005 (2009).

\bibitem{key}  Jouguet P.,  Kunz-Jacques S. \&Leverrier A. \textit{Phys. Rev. A} \textbf{ 84}, 062317 (2011)
\bibitem{recon} Bloch, M. , Thangaraj, A. , McLaughlin, S.W. \& Merolla, J.M. LDPC-based secret key agreement over the Gaussian wiretap channel. \textit{Proc. IEEE Information Theory Workshop} 1179-1183  (2006).

\bibitem{epr} Reid, M. D. et al., Colloquium: The Einstein-Podolsky-Rosen paradox: From concepts to applications. \textit{ Rev. Mod. Phys.} \textbf{81},  1727-1751 (2009).

\bibitem{extremality} Wolf, M. M. , Giedke, G. \&  Cirac J. I. Extremality of Gaussian quantum states.  \textit{Phys. Rev. Lett.}  \textbf{96} 080502 (2006).
\bibitem{proof1} Garc\'ia-Patron R. \& Cerf, N. J. Unconditional optimality of Gaussian attacks against continuous-variable quantum key distribution. \textit{Phys. Rev. Lett.}  \textbf{97} 190503 (2006).
\bibitem{proof2} Navascues, M. , Grosshans,  F. \&  Ac\'in, A. Optimality of Gaussian attacks in continuous-variable quantum cryptography. \textit{Phys. Rev. Lett.}  \textbf{97}, 190502 (2006).
\bibitem{finetti} Renner, R. \&  Cirac, J. I. de Finetti representation theorem for infinite-dimensional quantum systems and applications to quantum cryptography. \textit{Phys. Rev. Lett.} \textbf{102}, 110504 (2009).


\bibitem{equiv} Grosshans, F. , Cerf, N. J. , Wenger, J. , Tualle-Brouri, R. \&  Grangier P. Virtual entanglement and reconciliation protocols for quantum cryptography with continuous variables. \textit{Quantum Inf. Comput.} \textbf{ 3},  535-552 (2003).
\bibitem{fred} Shen, Y. , Peng, X. , Yang, J. , \& Guo H. Continuous-variable quantum key distribution with Gaussian source noise. \textit{Phys. Rev. A} \textbf{83}, 052304 (2011).
\bibitem{vlad1} Usenko, V. C. \&  Filip, R. Feasibility of continuous-variable quantum key distribution with noisy coherent states. \textit{Phys. Rev. A} \textbf{ 81}, 022318 (2010). 
\bibitem{braunstein2005} Braunstein, S. L. Squeezing as an irreducible resource. \textit{Phys. Rev. A} \textbf{71}, 055801 (2005).
\bibitem{beta} Jouguet, P., Kunz-Jacques, S., \& Leverrier, A. Long-distance continuous-variable quantum key distribution with a Gaussian modulation. \textit{Phys. Rev. A.} \textbf{84} 062317 (2011).
\bibitem{our} Usenko, V. C. \&  Filip, R. Squeezed state quantum key distribution upon imperfect reconciliation. \textit{New J. Phys.} \textbf{13} 113007 (2011).
\bibitem{fiber} Dong, R. et al., Experimental evidence for Raman-induced limits to efficient squeezing in optical fibers. \textit{Optics Letters}, Vol. \textbf{33}, Issue 2, pp. 116-118 (2008).   


\end{thebibliography}
\end{document}